\documentstyle[amssymb,aps,prb,floats,twocolumn,epsfig]{revtex}

\begin{document}

\wideabs{ \title{Magneto-photoluminescence of GaN/AlGaN quantum wells:
\\ valence band re-ordering and excitonic binding energies}

\author{P. A. Shields, R. J. Nicholas} \address{Department of Physics,
Oxford University, \\Clarendon Laboratory, Parks Rd., Oxford, OX1 3PU,
U.K.}  \author{N. Grandjean, J. Massies} \address{CNRS, Centre de
Recherche sur l'H$\acute e$t$\acute e$ro-Epitaxie et ses Applications, Valbonne,
F-06560, France.}  
\date{\today } 
\maketitle

\begin{abstract}

A re-ordered valence band in GaN/AlGaN quantum wells with respect to
GaN epilayers has been found as a result of the observation of an
enhanced g-factor $(g^{*}\sim3)$ in magneto-luminescence spectra in
fields up to 55 T. This has been caused by a reversal of the states in
the strained AlGaN barriers thus giving different barrier heights for
the different quantum well hole states. From $k.p$ calculations in the
quasi-cubic approximation, a change in the valence-band ordering will
account for the observed values for the g-factors.

We have also observed the well-width dependence of the in-plane extent
of the excitonic wavefunction from which we infer an increase in the
exciton binding energy with the reduction of the well width in general
agreement with theoretical calculations of Bigenwald et al (phys. stat. sol. 
(b) 216, 371 (1999)) that uses a
variational approach in the envelope function formalism that includes
the effect of the electric field in the wells.

\end{abstract}

 }

\section{Introduction}

The recent understanding of the influence of macroscopic polarisation
in nitride materials has  inspired a closer look at the
properties of AlGaN-based quantum wells where these effects are
stronger than found with the more commercially attractive InGaN
system.

The typical optical characterisation of a GaN/AlGaN quantum well
sample finds a single peak in luminescence at energies that are
strongly dependent on the width of the wells and the aluminium content
of the barriers. The energies of these peaks have been quantitatively
understood in terms of the quantum confined stark effect (QCSE) from
which the electric field caused by the polarisation can be deduced. It
has been found that the field has a strong linear dependence on the
aluminium concentration reaching almost 1.5 MV/cm for an Al
composition of 27\% \cite{Grandjean19994,Grandjean199919}.

The optical quality, as determined from the emission efficiency, is
better in the InGaN/GaN system compared to the GaN/AlGaN structures,
whereas for the emission linewidth the opposite is true. The latter is
likely to be related to the position of the ternary alloy in the
structure, it being in the active region in InGaN structures where an
inhomogeneous composition would have a greater effect. Typically the
best linewidths in InGaN/GaN structures are about 40 meV
\cite{Wang19983}, which can be compared with about 20 meV for
GaN/AlGaN wells \cite{Grandjean19994}. The development of these high
quality AlGaN/GaN QW's has now allowed
measurements to be made of the magneto-optical properties in nitride
based quantum wells.

Due to the linewidths already mentioned, the large effective masses
and the large exciton binding energies in these materials, it is necessary to use very high
magnetic fields to see any effects. This leads us to use pulsed-field magnets that can currently
give fields of up to 55 T. 

\section{Experiment}

We have performed magneto-optical experiments in both steady and
pulsed magnetic fields on single quantum well samples with different
widths of 4, 8, 12, and 16 monolayers (ML). One monolayer corresponds
to $2.59\AA$ therefore giving about 10, 20, 30 and 40$\AA$
respectively. The three samples described in this paper are
described in Table \ref{tab:sampledetails}. The structures were grown by molecular 
beam epitaxy on GaN templates on sapphire substrates with 200nm of AlGaN 
grown as a buffer layer before the quantum wells. For samples N298
and N307, the technique of lateral overgrowth on patterned GaN (ELOG) 
was used to improve the template quality\cite{Beaumont}. 
The sample widths and the aluminium composition
were determined through RHEED (reflection high energy electron
diffraction) spectra observed during the growth\cite{Grandjean199725}.

\begin{table} 
\begin{tabular}{cc}
Sample & Description\\ \hline
N257 & 4 SQWs of 4, 8, 12, 16 MLs separated\\ & by 100$\AA$ of AlGaN at 13\%. MBE template.\\
N307 & SQW of 8ML wide with AlGaN \\ & barriers with 8\% Al. ELOG template.\\ 
N298 & 4 SQWs of 4, 8, 12, 16 MLs separated\\ & by 100$\AA$ of AlGaN at 8\%. ELOG template.\\
\end{tabular}
\vspace{0.2cm}
\caption{Details of the samples grown by MBE.}
\label{tab:sampledetails}
\end{table}

The results that will be discussed are low and high excitation
photoluminescence (PL) with fields up to \mbox{55 T} at 4.2 K where the
effects of the field can be seen through a shift of the luminescence
peaks.

For the high excitation density PL the source was a pulsed (5ns)
frequency-quadrupled Nd:YAG laser at 266nm, whereas for the low
excitation density it was the 244nm line from a CW frequency-doubled
Argon laser. For the latter in the pulsed fields, a masked chopper
wheel was used to produce a train of short laser pulses, $\sim$0.3ms
duration, at $\sim$100ms intervals. A pulse from this train triggered
the magnet system so that the next laser pulse coincided with the
maximum of the pulsed field, where there is little change in the field
($\leq$1\%) over a 0.3 ms timescale. A CCD cooled to $-70^{o}C$ attached to a
quarter metre spectrometer was used to detect the PL.

\section{Results}

\begin{figure}
\epsfxsize=85mm
\epsffile{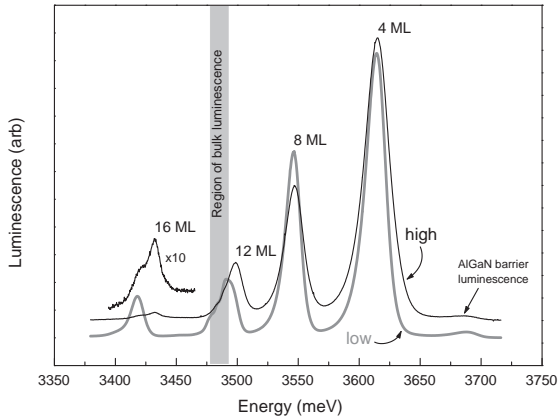}
\vspace{0.2cm}
\caption{High and low excitation photoluminescence of sample N257
from four separate quantum wells, of widths 4, 8, 12, 16 monolayers
(10, 20, 30, 40 $\AA$) scaled for a comparison of the peak positions.}
\label{fig:highlow}
\end{figure}

Figure \ref{fig:highlow} shows the low and high excitation PL for sample N257. Several
PL peaks can be seen that can be attributed to the different
wells. Between the low and high excitation, we have observed a shift
in the PL that is strongly dependent on the width of the wells,
ranging from 1meV for the narrowest up to 20meV for the widest. This
is in agreement with the idea that in-built electric fields play an
important role in this sample via the quantum confined Stark effect
\cite{Bernardini199718} and that they have been screened by the
excited carrier population. Previous measurements have calculated the
field to be ~550 kV/cm \cite{Grandjean19994}. The most obvious
consequence of this is seen through the observation of PL from the
widest well at 3.42 eV, which is below the band gap energy of bulk GaN
at 3.49 eV. Unfortunately, for this sample, the bulk feature coincides with the PL
from the 12 ML QW, so that luminescence from the bulk template
superimposes a structure onto this peak. 

The low intensity excitation is of the order $0.25 W cm^{-2}$ and 
can be considered to be in the regime where self-screening of the 
electric field is negligible. The exact excitation intensity of the 
pulsed laser is difficult to determine presisely, but can be estimated 
to be greater by about five orders of magnitude.
The magnetic field results discussed below were therefore taken using 
the low intensity quasi-CW excitation. 

\begin{figure}
\epsfxsize=85mm
\epsffile{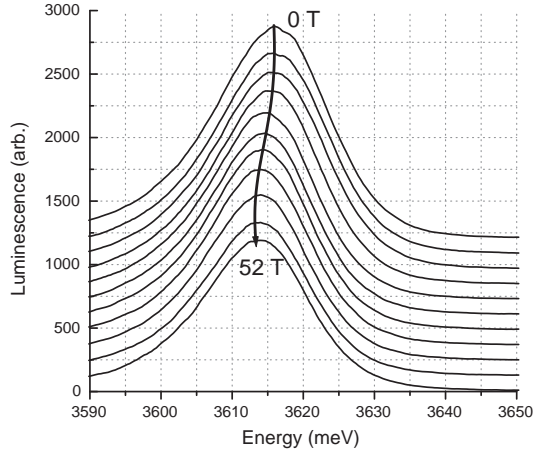}
\vspace{0.2cm}
\caption{Figure 2: Magnetic field dependence of the low excitation
photoluminescence for the 4 monolayer quantum well in sample N257. The
spectra are offset for clarity.}
\label{fig:n257spectra}
\end{figure}

\begin{figure}
\epsfxsize=85mm
\epsffile{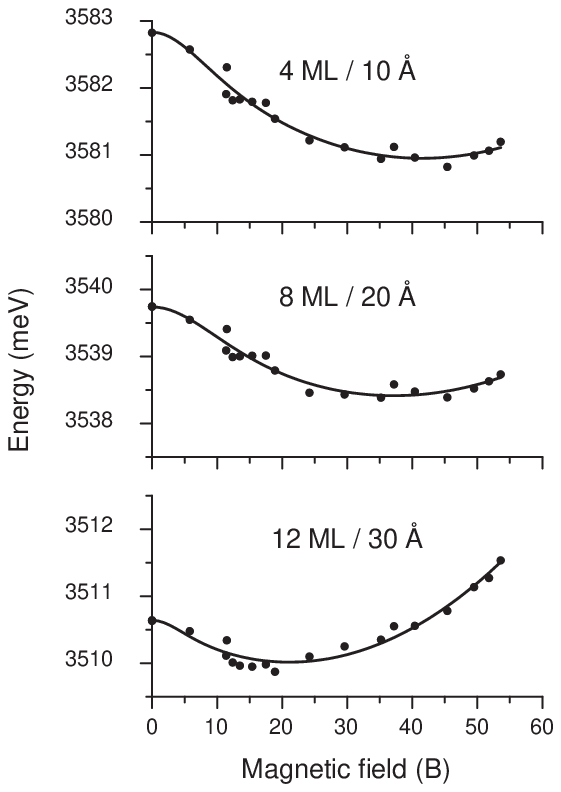}
\epsfxsize=85mm
\epsffile{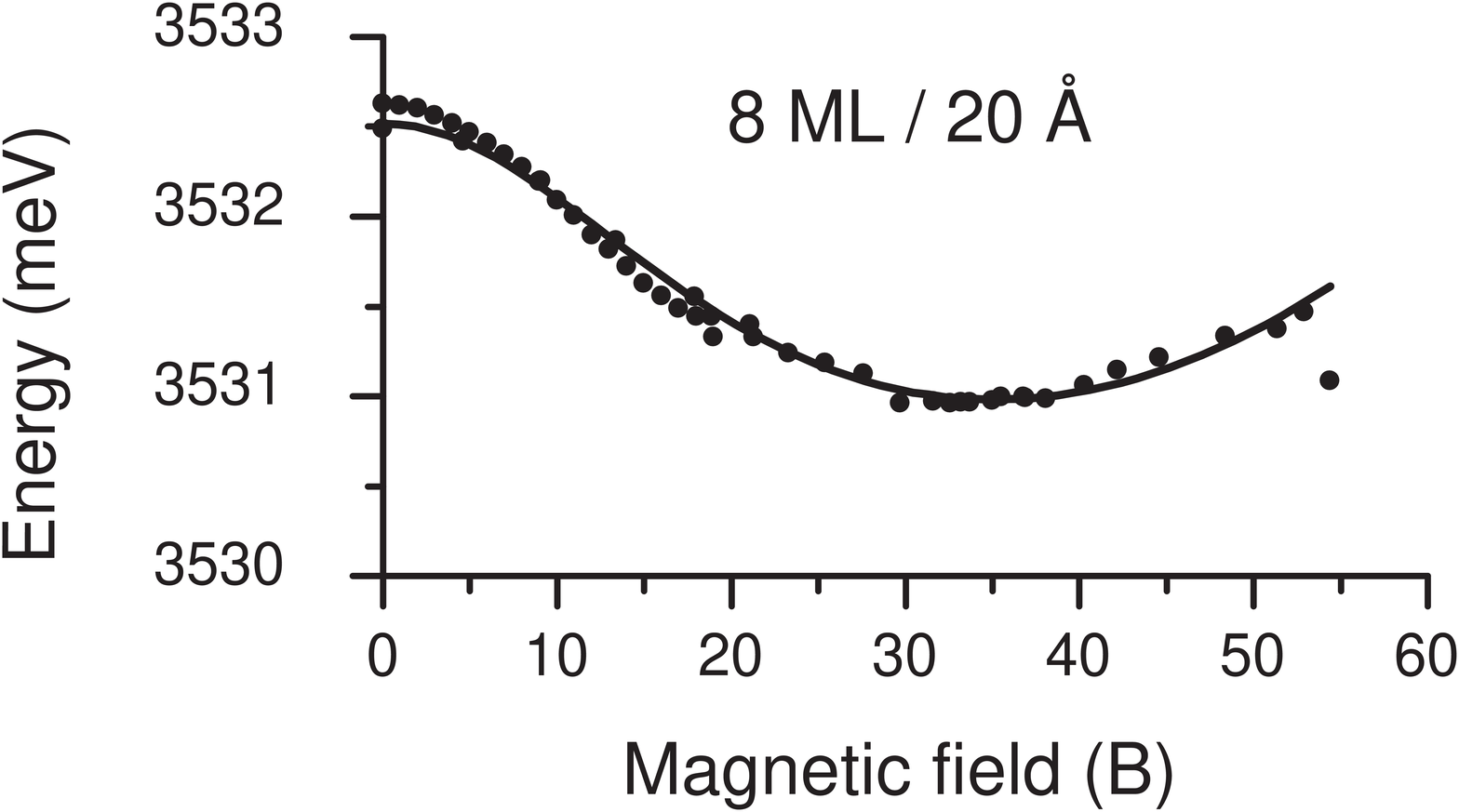}
\vspace{0.2cm}
\caption{Magnetic field dependence of the low excitation
photoluminescence for the different quantum wells for sample N298
(top) and N307 (bottom), along with the fits to the data as described
in the text.}
\label{fig:shifts}
\end{figure}

The effect of the magnetic field on the 4ML QW in sample N257 can be
seen in Figure \ref{fig:n257spectra}, and shows a shift of the luminescence to lower
energy. The form of this shift is dependent on the width of the wells,
and is shown in Figure \ref{fig:shifts}. In order to determine the peak positions a
\emph{centre of mass} method was used in preference to a peak fitting
routine \cite{Kukushkin199217}. This decision resulted from the
difficulty of deciding the correct form of the peak as it is slightly
asymmetric and thus causes problems for a Gaussian fit. The centre of mass is 
also known as the first moment, and is given by,

\begin{equation}
\label{eq:com}
M_1 = \frac{{\int {I(E)EdE} }}{{\int {I(E)dE} }}.
\end{equation}

For the widest wells there is a diamagnetic blueshift of the
transition energy, but as the well width decreases we observe a
transition to a redshift, which for the 4ML well is approximately 2meV
at 50T. Eventually at the highest field the redshift stops and appears
to be overtaken by a smaller diamagnetic term.

For the widest well, two components are observed that have different
dependencies on magnetic field. The observation of two peaks has
previously been suggested to be due to fluctuations in the well width,
corresponding to 16 and 17 ML's. We see an enhancement of the 16 ML
peak with field so that it dominates the spectra at high field, which is consistent
with this proposal and the suppression of the in-plane transport. The
excitons are then less able to find the potential minima caused by the
fluctuations before recombining.

The shifts for all the wells can be described by the expression, $E(B)
= E_0\pm{\textstyle{1\over2}}g^*\mu_BB+\gamma_2B^2$ so that the
overall dependence for each Zeeman-split component is determined by
the relative magnitudes of the linear Zeeman and quadratic diamagnetic
terms. The linear Zeeman shift causes the transition to split into
two, of which we see predominantly the lowest energy component. The
S-shape observed especially for the narrower wells at the lower fields
can be explained as resulting from a significant thermal population in
the upper component whilst the Zeeman splitting is still small. This
effect is particularly noticeable when the width of the luminescence
is large compared with the shift in magnetic field and $kT$, as is the
case in these samples.

The magnetic field dependence of the centre of mass can then be
expressed in terms of the populations in each component,
$n\uparrow$ and $n\downarrow$,

\begin{equation}
\label{eq:shiftcom}
E=E_0+{\textstyle{1\over2}}g^*\mu_B B\left({\frac{{n\uparrow-n\downarrow }}{{n \uparrow + n \downarrow }}}
\right) + \gamma _2 B^2 ,
\end{equation}
	
with the ratio, $\frac{{n \uparrow }}{{n \downarrow }}$, given by a
Boltzmann distribution. Table \ref{tab:fitting} shows the values deduced from the
experiments for the characteristic temperatures, g-factors and
diamagnetic shift coefficient. These last two parameters will be
individually discussed in the following sections. The temperatures
deduced simply reflect a heating of the carriers by the high
excitation intensities onto the sample. 
These were necessarily rather high in order to detect sufficient signal
in the short time duration of a single pulse ($\sim$0.3 ms).

\begin{table}
\begin{tabular}{cccc}
Sample N257 & T (K) & g* & $\gamma_2 (\mu eV/T^2)$\\
4 ML QW & 15(1) & 3.1 (1) & 0.99 (7)\\
8 ML QW & 13(1) & 2.9 (2) & 1.3 (1)\\
12 ML QW & 13(1) & 3.3 (3) & 2.1 (2)\\
16 ML QW & 5(5) & 1.8 (4) & 3.0 (3)\\
\hline
N298 &  &  & \\
4 ML QW & 9(3) & 3.1 (2) & 1.1 (1)\\
8 ML QW & 8(3) & 2.5 (2) & 1.0 (1)\\
12 ML QW & 5(5) & 2.0 (2) & 1.4 (1)\\
\hline
N307 &  &  & \\
8 ML QW & 17(3) & 3.4 (2) & 1.5 (1)\\
\hline
G889 &  &  & \\
Bulk & - & 0 & 2.04 (3)\\
\end{tabular}
\vspace{0.2cm}
\caption{The magnetic field fitting parameters for the different
GaN/AlGaN samples along with a bulk reference.}
\label{tab:fitting}
\end{table}

\section{Discussion}

\subsection{Zeeman splitting}

\begin{figure}
\epsfxsize=3 in
\epsffile{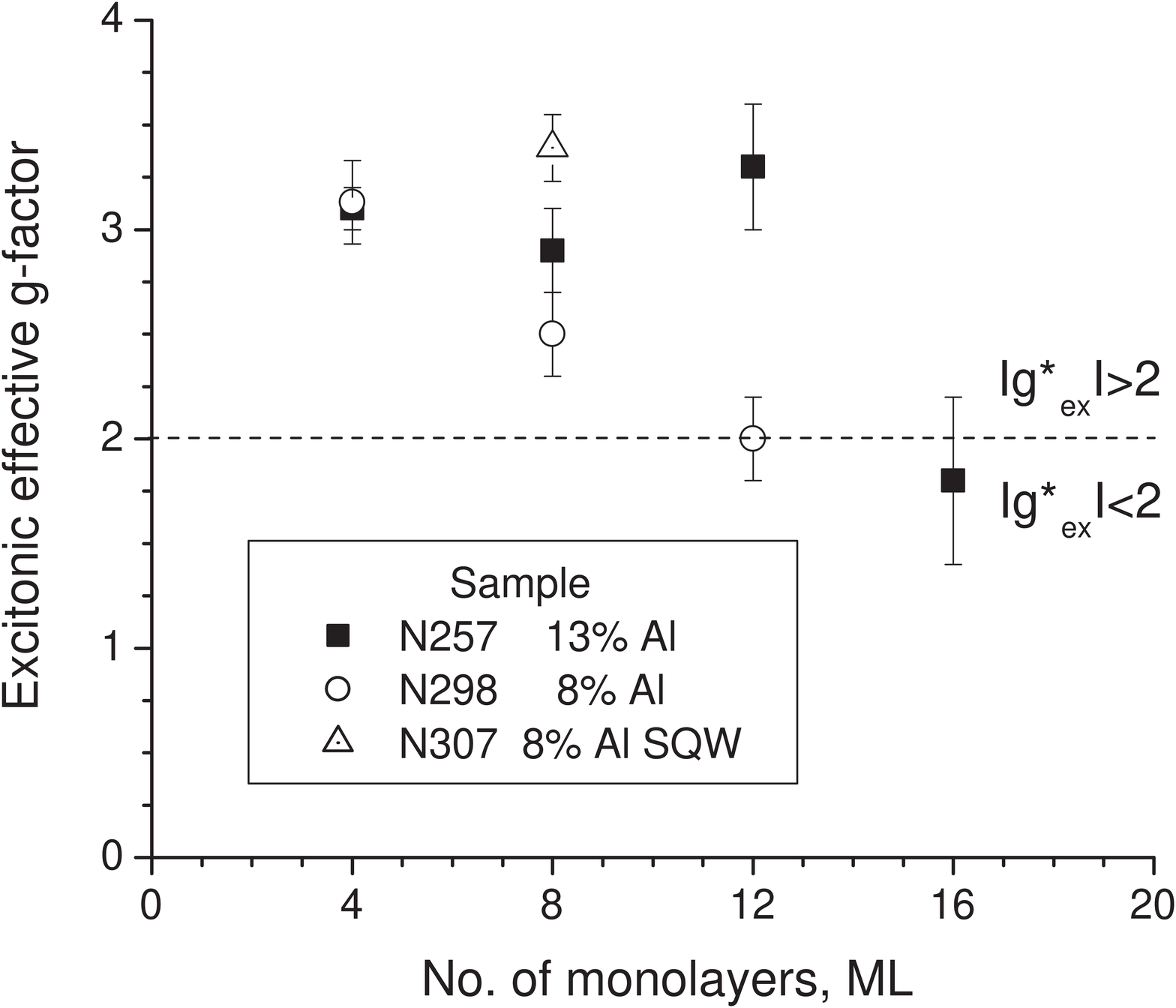}
\vspace{0.2cm}
\caption{The experimentally determined well width dependence of the
g-factors.}
\label{fig:gfactors}
\end{figure}

The g-factors for the different samples are compared in Figure \ref{fig:gfactors}, and
show a general trend towards $g^*\sim3$ for the narrower wells, with it
reducing to $g^*\sim2$ for the wider wells. This is particularly
noticeable in sample N298.

The observation of a Zeeman splitting through a clear redshift is
surprising, since in the magneto-reflectivity of GaN epilayers, no
such splitting is seen. From the results in bulk GaN the lowest
transition, associated with the A valence band, does not split
for the same orientation as the experiments described here
\cite{Shields19995,Stepniewski19996}. This is due to a
cancellation of the electron and hole g-factors when the magnetic
field is parallel to the c-axis. Even for the B valence band there is
a similar compensation leading to an excitonic g-factor, $g^*=1.24$. So
as both theoretical and experimental work on AlGaN heterojunctions
\cite{Knap19997} suggest that the electron effective g-factor should
not change from a value of $\sim2$ as a result of confinement, we conclude
that instead the hole g-factor must have drastically changed in the
narrower quantum wells due possibly to a significant change in the
valence band structure.

The reduced symmetry of the wurtzite with respect to the cubic
structure is represented in the Hamiltonian through the crystal
field. If spin-orbit effects are neglected, the crystal field reduces
the degeneracy of the valence bands with different z-components of
orbital angular momentum. With spin-orbit terms included there are
then the three usual bands; A, B, and C with symmetries
$\Gamma_9$, $\Gamma_7$, and $\Gamma_7$ respectively, that have the following
separations,

\begin{equation}
\label{eq:crystalfield}
E_A\!-\!E_{B\!,C}=\!
\frac{1}{2}\!\left\{\!\left(\!\Delta_{cr}\!\!+\!\!\Delta_{so}\!\right)\!
\mp\!\left[\left(\! \Delta_{cr}\!\!+\!\!\Delta_{so}\!\right)^2\! -
\!\frac{8}{3}\Delta_{cr}\Delta_{so}\! \right]^\frac{1}{2}\! \right\},
\end{equation}
	
where $\Delta _{cr} $ and $\Delta _{so} $ are the crystal field
and spin-orbit energy terms respectively \cite{Bir197422}.

This is for the case of a bulk semiconductor. With the effects of a
two-dimensional quantum confinement, the energies are usually
understood in terms of the envelope function approximation that are
then dependent on: the effective masses in the well and barrier, the
valence band potential, and the width of the well. This causes the
energies of the bands to become unrelated to the bulk energy terms in
Equation \ref{eq:crystalfield}. Instead effective spin-orbit and crystal field terms can be
used to account for the splittings of the A, B and C valence bands,
where the nomenclature is now strictly in terms of the band symmetries
and not the ordering.

\begin{figure}
\epsfxsize=85mm
\epsffile{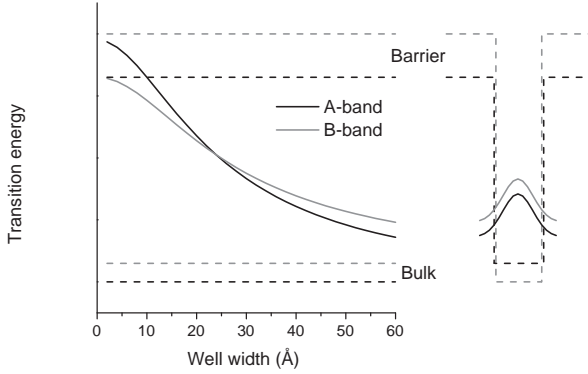}
\vspace{0.2cm}
\caption{A schematic diagram based on a finite quantum well model
showing how the quantum well energies can cross at a particular well
width for two bands with different offsets.}
\label{fig:qwschematic}
\end{figure}

For the GaN/AlGaN system, it is known that in bulk AlN the crystal
field splitting has the opposite sign to that of GaN thus inverting
the order of the states. This would lead to a reversal of the states
at some critical value of $x$ in $Al_xGa_{1-x}N$ 
\cite{Kim19978}, though this
value is not experimentally known. The more important consequence for
this work is that this will cause different valence band offsets for
the different bands and Figure \ref{fig:qwschematic} shows how this effect can lead to a
reversal of the quantum well states, even when the same masses in the
z-direction, $m_\parallel $, are used. (In fact $m_\parallel$ is
identical for the A and B valence bands.) This calculation was carried
out for a symmetrical quantum well, which is known to be invalid for
these samples. However the effect of a strong electric field will be
to bring the confinement energy closer to the barrier offsets thus
making them more important and enhancing the reversal effect.

\begin{figure}
\epsfxsize=85mm
\epsffile{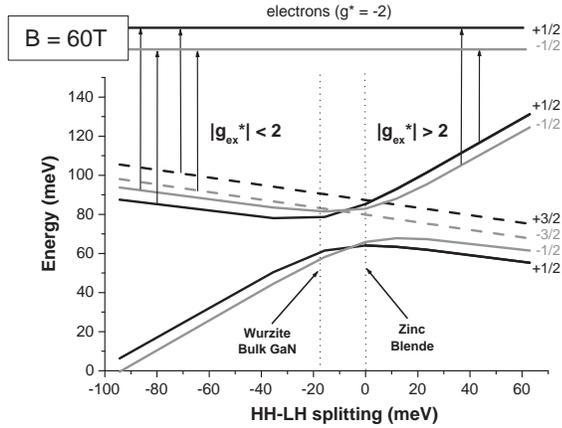}
\vspace{0.2cm}
\caption{$k.p$ results for GaN/AlGaN showing how the crystal field
splitting of the valence band affects the spin-ordering of the valence
band states at 60 T. $\Gamma_7$ states are indicated as solid lines, and $\Gamma_9$
states are dashed. The vertical arrows show the allowed optical
transitions.}
\label{fig:kp}
\end{figure}

A reversal of states has a profound effect on the spin-ordering of the
valence band that can be clearly seen through the k.p calculations of
Figure \ref{fig:kp} within the quasi-cubic model at a magnetic field of \mbox{60 T},
using the band parameters deduced by Stepniewski et al\cite{Stepniewski19996}. With this model, the wurtzite
Hamiltonian can be approximated as a cubic Hamiltonian that is
uniaxially strained along the [111] direction. The \emph{crystal field} from
the wurtzite vocabulary can then be translated into a heavy-light hole
splitting (HH-LH) in cubic language. Therefore with zero strain,
corresponding to zero HH-LH splitting or zero crystal field, we find
the usual zinc-blende band and spin ordering. The bulk wurtzite case
is also indicated, corresponding to a finite strain or crystal field,
and this correctly reproduces the spin-ordering in GaN
\cite{Stepniewski19996}, whereas AlN would correspond to the far right
of the graph with a large positive HH-LH splitting. The crossing, or
more correctly anti-crossing, of the $\Gamma_7$ states causes a reversal of
the spin-ordering thus changing the sign of the g-factor for the
lowest\footnote{Lowest in terms of the hole energy.} $\Gamma_7$ hole band. 
This process also lowers its energy below the $\Gamma_9$
A-valence band so that transitions involving this band are favoured in
luminescence. Excitonic g-factors less than two occur for the lowest
bands to the left of the anti-crossing in Figure \ref{fig:kp} as indicated,
whereas they are greater than two to the right.

For a GaN/AlGaN quantum well structure the precise band ordering of
the confined states is hard to predict since it is critically
dependent on both the well and barrier properties due to the electric
field. As an extra tool, X-ray measurements can be used to examine the
crystal properties and give information on the strain that is
present. In these samples, it has been shown that the AlGaN barriers
have the same lattice constant as the GaN template, it being larger
than if they were unstrained, resulting in them having a tensile
biaxial strain \cite{Grandjean199926}. Therefore the well material is
unstrained and has a normal wurtzite GaN valence band structure,
whilst the barrier has the opposite ordering due to
the AlN contribution in the alloy and the strain present.

The latter can be estimated from using a linear relationship for the
lattice constant, giving an in-plane strain, $\varepsilon _{xx} =
\varepsilon _{yy} = 0.31\%$ for an aluminium content of 13\%. This
contributes to the crystal field splitting through $\Delta _{cr} +
{\textstyle{3 \over 2}}D_3 \varepsilon _{zz}$ with $\varepsilon _{zz}
= - {\textstyle{{2C_{13} } \over {C_{33} }}}\varepsilon _{xx} \approx
- 0.51\varepsilon _{xx}$, and $D_3 \sim 6eV$ to give ${\textstyle{3 \over 2}}D_3 \varepsilon _{zz}=
14meV$ \cite{Kim19978}. As experimental values of $\Delta _{cr}$ for
GaN are very close to this value (11-15 meV\footnote{11,12meV from review 
in Kim et al\cite{Kim19978}, $14.9\pm0.3 meV$ from reflectivity of sample G889}),
 once the contribution
of the AlN is also included, the barriers are very much in the regime
for a reversal of states to occur. Figure 7 of Kim et al\cite{Kim19978}
shows a calculation for a similar tensile biaxial strain in GaN that
clearly shows a reversal of the states with the anti-crossing also
giving the lowest valence band a heavier mass than would be otherwise
expected.

Therefore our results suggest that the ordering of the valence band
has changed for narrow quantum wells, perhaps as a result of the lower
barrier height for the crystal field $\Gamma_7$ band caused by the strain and
the aluminium content in the barriers. The reordering of these states
will cause the ground state to have a $\Gamma_7$ character and the inverted
spin splitting would account for the observed enhanced effective
g-factor.

The increase of the value for sample N307 in Figure \ref{fig:gfactors} with respect to
the same width in the other samples can be understood in terms of the
distributions of the electric fields. Where more than one quantum well
is present, the finite thickness of the barrier allows a
redistribution of the polarisation fields between both the well and
the barrier, whereas the field is solely in the wells for the
SQW. This will push the hole states further into the barrier material.

\section{Diamagnetic shift}

The magnitude of the diamagnetic shift in equation \ref{eq:shiftcom} is determined by
the size of the exciton wavefunction in the plane perpendicular to the
field. From this it is possible to infer the excitonic binding energy
and the following section describes how these parameters have been
determined for the different quantum wells along with values
corresponding to the bulk case.

The effect of a magnetic field can be considered as an extra confining
potential on the already localised excitonic system in both 2D and 3D
cases. The bulk data can be fitted very well to a full numerical
calculation for a 3-dimensional hydrogen atom in a magnetic field
\cite{Makado19869} that uses the excitonic binding energy and the
reduced mass as fitting parameters. The binding energy for our
particular sample was determined by Neu et al\cite{Neu199910} as 24.1 meV
by identifying the 2s excited state, and this enables the fitting to
give a reduced mass of 0.180(2) $m_e$ as the only free parameter. This
numerical calculation reduces to a $B^2$ dependence in the low field
limit, so that by fitting the low field data we obtain a coefficient,
$\gamma _{\rm 2} {\rm = 2}{\rm .04(3) }\mu eV/T^2$ that we can use in
a direct comparison between the bulk and QW data.

The diamagnetic coefficient for a quantum well is given by
\cite{Walck199811},

\begin{equation}
\label{eq:walck}
\gamma_2=\frac{e^2}{8\mu}\left\langle \rho^2 \right\rangle,
\end{equation}
	
where $\mu$ is the reduced mass of the exciton and $\rho$ is the separation of
the electron and the hole in the plane of the quantum well. Therefore
the diamagnetic coefficient contains information about the zero-field
properties of the exciton and can give a value for the in-plane extent
of the wavefunction.

From the 1s orbital in the hydrogen atom, the expectation value of $r^2$
is, $\left\langle {r^2 } \right\rangle = 3{a_0^*} ^2 $, where $a_0^* $
is the effective Bohr radius.  $r^2$ can then be converted to an in-plane
size through $\left\langle {\rho ^2 } \right\rangle = {\textstyle{2 \over
3}}\left\langle {r^2 } \right\rangle $. If the exciton is assumed to remain essentially three
dimensional, the effective binding energy, $R^*$, can then be deduced
through scaling with respect to the hydrogen atom,

\begin{equation}
R^* = \frac{1}{{\varepsilon _r }}\frac{{a_0^H }}{{a_0^* }}R^H ,
\end{equation}

where $a_0^H$ is the Bohr radius and $R^H$ is the Rydberg
constant. This gives the following expression,

\begin{equation}
R^*  = \sqrt {\frac{{e^2 }}{{4\mu }}\frac{1}{{\gamma _2 }}} \frac{{a_0^H R^H }}{{\varepsilon _r }}.
\end{equation}

\begin{table} 
\begin{tabular}{cccc}
Sample & Width,($\AA$) & In-plane extent  & Exciton binding \\
N257 &  (approx)& of exciton, ($\AA$) & energy, (meV)\\
4 ML & 10 & 28.5 (1.0) & 36.5 (1.3)\\ 
8 ML & 20 & 33.0 (1.2) & 31.5 (1.2)\\ 
12 ML & 30 & 41.7 (2.0) & 24.9 (1.2)\\ 
16 ML & 40 & 49.4 (2.5) & 21.0 (1.1)\\
\hline
N298 &  &  & \\
4 ML & 10 & 30.0 (1.4) & 34.6 (1.6)\\ 
8 ML & 20 & 28.3 (1.7) & 36.6 (2.2)\\ 
12 ML & 30 & 33.9 (1.2) & 30.7 (1.1)\\
\hline
N307 &  &  & \\
8 ML  & 20 & 35.0 (1.2) & 39.6 (1.0)\\
\hline
G889 &  &  & \\
Bulk & - & 40.9 (5) & 25.4 (2)\\
\end{tabular}
\vspace{0.2cm}
\caption{The in-plane extent and exciton binding energy for the different
quantum well widths deduced from the diamagnetic shift of the luminescence.}
\label{tab:diagshift}
\end{table}

The results from this calculation have been summarised in Table \ref{tab:diagshift} ($\varepsilon_r
= 9.8$ \footnote{3D average; $\varepsilon=(\varepsilon_\bot)^\frac{2}{3} (\varepsilon_\parallel)^\frac{1}{3}$} 
\cite{Barker197312}, $\mu=0.18m_0$ \footnote{This value is taken from the bulk calculations and is used as a
first approximation. The actual value will depend on the precise band
ordering and mixing effects as described in the section entitled Zeeman splitting.}). The similarity of the
binding energy for the bulk deduced in this way, 25.3 meV compared to
24.1 meV from the 1s-2s separation, illustrates the accuracy of this
technique.

\begin{figure}
\epsfxsize=85mm
\epsffile{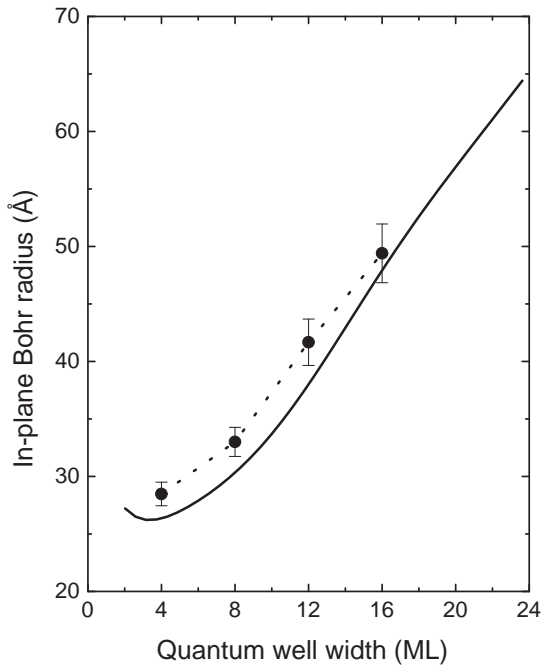}
\vspace{0.2cm}
\caption{Comparison between the experimental (points) and theoretical
values (solid line) for the in-plane extent of the excitonic
wavefunction for different well widths. The theoretical values are determined 
from considering a two-parameter variational trial function for a series of
GaN/AlGaN QW's with Al barrier composition x=0.27 
. The experimental values are from sample N257
and are deduced from the diamagnetic shift of the luminescence.}
\label{fig:inplane}
\end{figure}

\begin{figure}
\epsfxsize=85mm
\epsffile{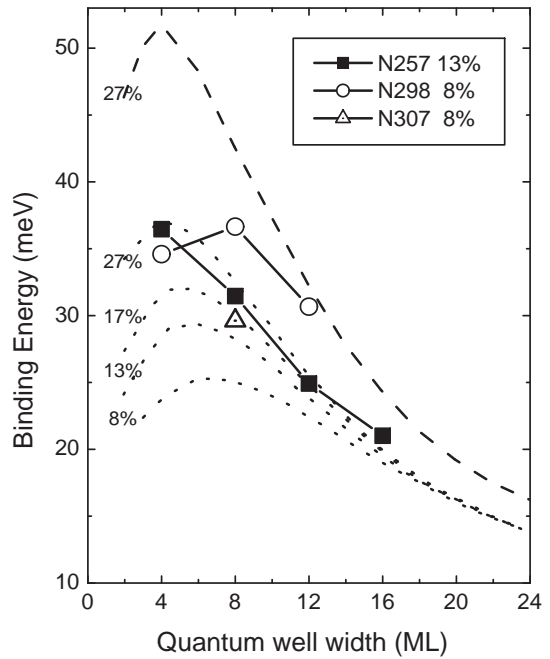}
\vspace{0.2cm}
\caption{Experimental values of the binding energies shown with calculations
 of the well width dependence of the exciton binding energies 
for a GaN/AlGaN SQW including internal electric fields. 
For a single parameter variational trial function (short dash), calculations are shown 
for four aluminium compositions, x= 0.27, 0.17,
0.13, 0.08, whereas for a two-parameter function (long dash), x=0.27.}
\label{fig:be}
\end{figure}

Table \ref{tab:diagshift} shows that as the well width is reduced, the exciton size is
also reduced resulting in a strong enhancement of the binding energy
for the narrow wells, in agreement with Bigenwald et al\cite{Bigenwald199913} and Grandjean et al\cite{Grandjean19994}. 
We also observe a reduction of the exciton
binding energy for the widest well due to the separation of the
electron and hole to the separate interfaces of the QW. Our results
can be compared with the calculations from Bigenwald et al\cite{Bigenwald199913}
shown in Figure \ref{fig:inplane} and Figure \ref{fig:be} of the in-plane extent and binding
energy of excitons in GaN/AlGaN quantum wells for different
compositions and well widths.

These variational calculations of the excitonic wavefunctions were
used to investigate the ground state of heavy-hole, free excitons
under different levels of approximation. Two different forms of the 
variational ground state wavefunction were considered in the calculations.
Both included a single variational parameter to govern the lateral extent 
of the wavefunction, however for one an additional variational parameter
was included in the determination of the separation of the electron and hole 
along the confinement axis. In the single parameter trial function this separation 
is solely determined by the electric fields present in the structure, whereas 
an additional contribution from the Coulomb attraction is allowed for 
in the two-parameter trial function.

The experimental results from N257 are shown alongside the theoretical calculations 
of the in-plane pseudo-Bohr radius in Figure \ref{fig:inplane}. 
Our results agree well with the theory despite the lower
aluminium concentration in the barriers, 13\% compared to the value of
27\% used in the calculations.

There are larger differences between the binding energies deduced from
this parameter, and this can be understood from the simplicity of our
model and the greater influence that the higher aluminium
concentration is expected to have through the increased electric field.
However the data favours the inclusion of the field
rather than its omission in square well calculations\cite{Bigenwald199913}.

The single parameter trial function underestimates the binding
energy for a given aluminium composition and the more accurate 
two-parameter function is shown only for an aluminium composition, x=0.27. From
comparing results for x=0.27 for both models, it seems that the two parameter function
for the appropriate composition, x=0.08, \& 0.13, would 
accurately account for the data.

It should be pointed out the theoretical results are for the
heavy-hole exciton, which we have shown is not the lowest state for these
samples. This should not be significant for the purposes of comparison
because the excitonic mass will be predominantly determined by the
electron mass.  The relevant hole mass would be an average over the
top of the valence band within an exciton binding energy from the band
edge, and is likely to be heavy as a result of the valence band mixing
associated with the band reordering. Both theory and experiment use
the heavy-hole exciton mass as a first approximation.

In converting from the diamagnetic coefficient, $\gamma_2$, to the binding energy
the three-dimensional hydrogen atom was used as a basis rather than a
two-dimensional atom. 
This approximation remains reasonable due to both the shrinkage of the in-plane
wavefunction caused by the increased binding and the considerable
leakage of the bound state wavefunction out of the wells into the
barriers as evidenced by both the change in the valence band ordering
and the calculated maximum in the excitonic binding energy at around 4
ML. This model is not as thorough as that used in the theoretical
calculations where along with the \emph{quasi-2D} wavefunction, they also
use a self-consistent algorithm.

The energy shifts that we have discussed are very small in comparison with
the linewidth. It is worth thinking of the reasons behind this
and the prospects for future experiments. In the field of nitride
semiconductors, it has become automatic to blame any sample
inadequacies on the lack of a suitably lattice-matched
substrate. However recent work on homoepitaxially-grown quantum well samples has
suggested that there is perhaps a fundamental limitation in the
linewidth, similar to the linewidths present in this work, 
that is caused by the random positions of the group III
elements in the ternary barrier combined with the large hole masses
\cite{Grandjean200014,Gallart200016}. The strong internal fields, 
through causing a strong separation of the electron and hole wavefunctions 
to the opposite interfaces of the quantum well, thus make the optical 
properties very sensitive to these barrier irregularities. 

\section{Conclusions}

This work has presented magneto-luminescence data in fields up to 55 T
for three different GaN/AlGaN quantum wells that shows a strong well
width dependence of the shift of the luminescence peaks. The high
pulsed fields were essential in the resolution of
the different dependencies, and even when the immense amount of work
on the sample growth in the future is considered with the expected
improvement in quality, it is unlikely that this would allow lower
continuous fields to be used due to the fundamental linewidth
limitations present.

Our data shows that by observing an enhanced g-factor for the narrow
wells, the valence band must have changed significantly compared with
bulk GaN. We have attributed this to a reordering of the valence band
states in the strained AlGaN barriers, giving different barrier
heights for the different quantum well hole states. We have also
observed an increase in the exciton binding energy with the reduction
of the well width in agreement with calculations using a variational
approach in the envelope function formalism that includes the effect
of the electric field in the wells. Better agreement is obtained when
the trial wavefunction includes a consideration of the
three-dimensional Coulomb potential.

\section{Acknowledgements}

We are grateful to the EPSRC (UK) for the support of this work and 
particularly the Lasers for Science Facility for the provision of the
frequency-doubled Argon laser. P.A.S. acknowledges financial support
from Sharp Laboratories of Europe Ltd.

\end{document}